\documentclass[sigconf]{acmart}
\usepackage{subfigure}
\usepackage{multirow}
\usepackage{bbding}
\usepackage{enumitem}

\AtBeginDocument{%
  \providecommand\BibTeX{{%
    \normalfont B\kern-0.5em{\scshape i\kern-0.25em b}\kern-0.8em\TeX}}}

\copyrightyear{2022}
\acmYear{2022}
\setcopyright{acmcopyright}\acmConference[CIKM '22]{Proceedings of the 31st ACM International Conference on Information and Knowledge Management}{October 17--21, 2022}{Atlanta, GA, USA}
\acmBooktitle{Proceedings of the 31st ACM International Conference on Information and Knowledge Management (CIKM '22), October 17--21, 2022, Atlanta, GA, USA} \acmPrice{15.00}
\acmDOI{10.1145/3511808.3557106}
\acmISBN{978-1-4503-9236-5/22/10}

\begin{document}

\title{KEEP: An Industrial Pre-Training Framework for Online Recommendation via Knowledge Extraction and Plugging} 




\author{Yujing Zhang} \authornotemark[1]
\email{jinghan.zyj@alibaba-inc.com}
\author{Zhangming Chan}
\email{zhangming.czm@alibaba-inc.com}
\authornote{Yujing Zhang and Zhangming Chan contributed equally to this work.}
\affiliation{%
  \institution{Alibaba Group}
  \city{Beijing}
  \country{People's Republic of China}
}


\author{Shuhao Xu}
\email{xsh20@mails.tsinghua.edu.cn}
\affiliation{%
  \institution{School of Software, \\ Tsinghua University}
  \city{Beijing}
  \country{People's Republic of China}
}

\author{\mbox{Weijie Bian}}
\author{Shuguang Han}
\authornote{Shuguang Han is the corresponding author.} 
\email{weijie.bwj@alibaba-inc.com}
\email{shuguang.sh@alibaba-inc.com}
\affiliation{%
  \institution{Alibaba Group}
  \city{Beijing}
  \country{People's Republic of China}
}


\author{Hongbo Deng}
\email{dhb167148@alibaba-inc.com}
\affiliation{%
  \institution{Alibaba Group}
  \city{Beijing}
  \country{People's Republic of China}
}

\author{Bo Zheng}
\email{bozheng@alibaba-inc.com}
\affiliation{%
  \institution{Alibaba Group}
  \city{Beijing}
  \country{People's Republic of China}
}

\renewcommand{\shortauthors}{Y. Zhang and Z. Chan, et al.} 
\renewcommand{\authors}{Yujing Zhang, Zhangming Chan, Shuhao Xu, Weijie Bian, Shuguang Han, Hongbo Deng, Bo Zheng} 
\renewcommand{\shorttitle}{KEEP: Knowledge Extraction and Plugging for Online Recommendation} 

\begin{abstract}
An industrial recommender system generally presents a hybrid list that contains results from multiple subsystems. In practice, each subsystem is optimized with its own feedback data to avoid the disturbance among different subsystems. However, we argue that such data usage may lead to sub-optimal online performance because of the \textit{data sparsity}. To alleviate this issue, we propose to extract knowledge from the \textit{super-domain} that contains web-scale and long-time impression data, and further assist the online recommendation task (downstream task). To this end, we propose a novel industrial \textbf{K}nowl\textbf{E}dge \textbf{E}xtraction and \textbf{P}lugging (\textbf{KEEP}) framework, which is a two-stage framework that consists of 1) a supervised pre-training knowledge extraction module on super-domain, and 2) a plug-in network that incorporates the extracted knowledge into the downstream model. This makes it friendly for incremental training of online recommendation. Moreover, we design an efficient empirical approach for KEEP and introduce our hands-on experience during the implementation of KEEP in a large-scale industrial system. Experiments conducted on two real-world datasets demonstrate that KEEP can achieve promising results. It is notable that KEEP has also been deployed on the display advertising system in Alibaba, bringing a lift of $+5.4\%$ CTR and $+4.7\%$ RPM. 
\end{abstract}

\ccsdesc[500]{Information systems~Online advertising}
\ccsdesc[500]{Information systems~Retrieval models and ranking}

\keywords{Online Recommendation, Pre-training, Knowledge Extraction, Knowledge Plugging} 

\maketitle

\section{Introduction}
\label{sec:intro}
Large-scale industrial recommender systems generally present a hybrid list of items in various product types. Take Taobao App for example, its homepage recommender may offer a list consisting of online products, short videos, or advertisements. In practice, items from different product types are often recommended by separate subsystems. For example, short videos are provided by an individual video recommender. Therefore, each subsystem often records its own user feedback data and trains a separate model to serve online. Such a straightforward architecture ensures that each subsystem focuses on its own optimization objective and avoids the disturbance among different subsystems. It is worth noting that with respect to the dynamic change of user interests, large-scale recommender systems often exhibit data distribution shift over time. To deal with this problem, the common practice is to train a model by only utilizing the most recent training data.

\begin{figure}[t]
    \centering
    \includegraphics[width=0.75\columnwidth]{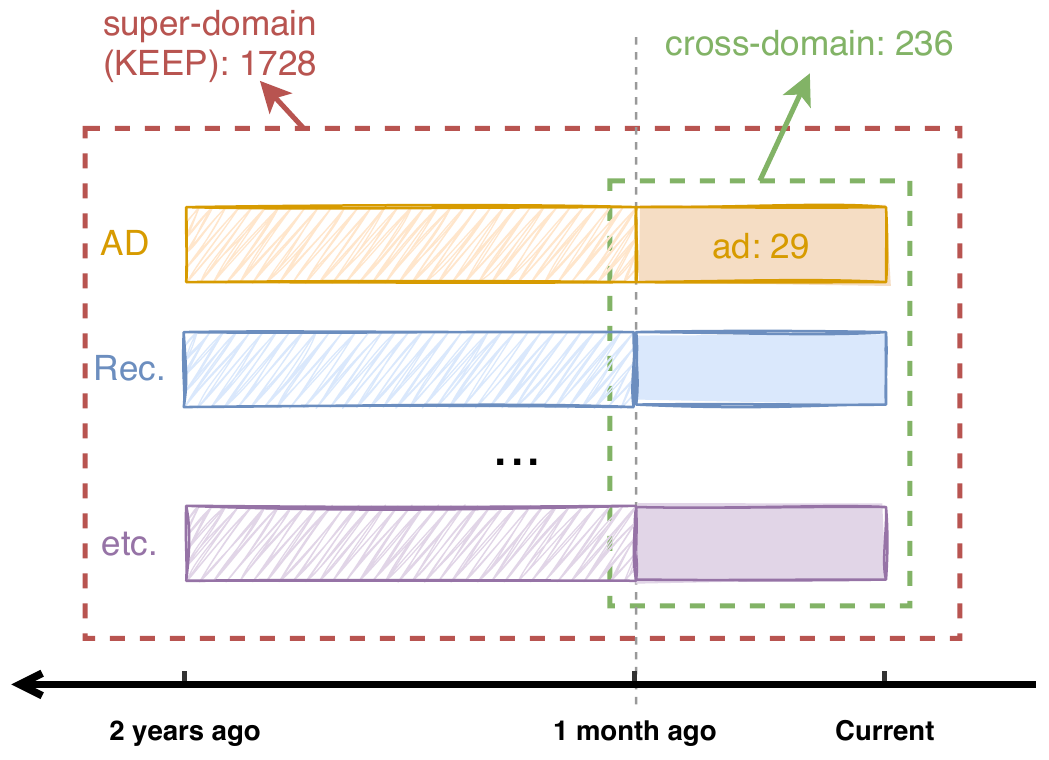} 
    \vspace{-3mm}
        \caption{The average number of impressions per user in Taobao homepage recommender. The ad indicates the average impression number per user in AD subsystem.}
    \label{fig:intro_case}
\end{figure}

We argue that such a data usage may lead to sub-optimal online performance because of the \textit{data sparsity} issue~\cite{li2015click,wu2021adversarial}. To illustrate this problem, we plot the average number of impressions for each user in the advertisement subsystem of Taobao homepage recommender. As shown in Figure ~\ref{fig:intro_case}, there are only an average of 29 impressions per user within one month. We argue that such a data volume is insufficient for model learning, particularly when considering the highly-skewed data distribution in industrial systems~\cite{wu2021adversarial}. To alleviate the \textit{data sparsity} issue, a series of works have experimented with the cross-domain modeling approach~\cite{erhan2009difficulty, sun2019bert4rec, zhou2020s3}, in which the knowledge from a richer source-domain is introduced to assist the model training of the sparser target-domain. Going back to Figure ~\ref{fig:intro_case}, by leveraging the data from other subsystems, the number of impressions for each user is increased to 236. 
The improvement from the cross-domain approach demonstrates the usefulness of utilizing data across different domains.

Herein, we attempt to introduce more data by extending the source-domain to a \textit{super-domain} which contains web-scale and long-time impression data. Now, in Figure ~\ref{fig:intro_case}, each subsystem has a total of 1728 impressions for each user available for model training. To be consistent with the naming of super-domain, we rename the subsystem as \textit{sub-domain}. In practice, a super-domain may contain tens of hundreds times of training samples comparing to a sub-domain. 
The excessively amount of data requires that the modeling of the super-domain should not hinder the consumption of training data in the sub-domain. However, conventional cross-domain models often adopted the coupled architecture, such as the share-bottom structure~\cite{hu2018conet}, making the super-domain model greatly affect the modeling of sub-domain. Hence, most cross-domain approaches are unsuitable for super-domain modeling. 

To handle the above shortcoming, we can decompose the modeling of super-domain and sub-domain in two separated stages, with the first stage focusing on extracting knowledge from the super-domain, and the second stage on utilizing the extracted knowledge. 
This workflow is close to the process of pre-training and fine-tuning. However, such an approach is inapplicable for online recommendation due to the need of frequent training. 
Specifically, to keep pace with online data distribution shift\footnote{The data distributions in online recommender system can experience shifts due to many factors such as seasonality and promotions.}, industrial recommender systems usually refresh their prediction models at daily or hourly basis. 
As we know, fine-tuning mechanism generally loads pre-trained parameters and conducts little tuning to avoid catastrophic forgetting~\cite{kirkpatrick2017overcoming}. Frequent training, as did in the online recommender systems, will accelerate the forgetting of pre-trained parameters. 

In this work, we propose a novel industrial \textbf{K}nowl\textbf{E}dge \textbf{E}xtraction and \textbf{P}lugging (\textbf{KEEP}) framework, to conduct the knowledge extraction from super-domain and further enhance the sub-domain modeling with extracted knowledge. KEEP consists of two stages.

In the first stage, we focus on the pre-training of a knowledge extraction module. Different from self-supervised pre-training approaches, we design a supervised pre-training for KEEP. This is mainly attributed to the simplicity for obtaining labeled feedback data in recommender systems. 
Specifically, we adopt the DIN architecture~\cite{zhou2018din} for the knowledge extraction module, and design three pre-training tasks including the predictions of click, purchase and add-to-cart. 
The extracted knowledge from the first stage may have various forms: user-level knowledge, item-level knowledge and user-item-interaction knowledge. 
In the second stage, we design a plug-in network which can effectively incorporate the extracted knowledge to downstream tasks.  Different from the fine-tuning mechanism, the plug-in network projects the extracted knowledge to a vector sharing the same shape as the output of the m-th dense layer of any downstream model.
Then, we incorporate the projected vector into the output of the m-th dense layer through add operation rather than common concatenate, so that it will not change downstream model architecture.
Based on such as plug-in network, we can easily adapt model parameters from a previously well-trained downstream model. 
This also brings several additional benefits such as friendly with incremental training and easily extensible once additional knowledge is available (details in Section~\ref{sec:kp}).

In order to deploy KEEP on an industrial system, and serve multi-scene and multi-task downstream models, we designed a separate service, called General Knowledge Center (GKC). GKC follows the decomposition and degeneration strategies to cache the extracted knowledge (from the first-stage of KEEP) in a high-performance parameter server. In GKC, knowledge extraction process is not calculated in real-time so that it only leads to a trivial increase of latency for downstream tasks, which is particularly important for industrial recommender systems serving massive traffic. GKC has been deployed in the display advertising system in Alibaba. In Section~\ref{sec:empirical}, we will introduce our hands-on experience for the implementation of KEEP in our system.

The main contributions of our work are summarized as follows:
\begin{itemize}[leftmargin=*]
\item In the context of online recommendation, we propose a novel KEEP framework to extract knowledge from super-domain, and design a plug-in network to incorporate the extracted knowledge into sub-domain downstream tasks. 
\item We design an efficient empirical approach for KEEP and introduce our experience during the implementation of KEEP in large-scale industrial recommender systems. 
\item KEEP has been deployed in the display advertising system in Alibaba, bringing a lift of $+5.4\%$ CTR and $+4.7\%$ RPM. 
\end{itemize}

\section{Related Work}


\label{sec:related}

\subsection{Cross-Domain Recommendation}

To handle the long-standing data sparsity problem, cross-domain recommendation (CDR) has been proposed to leverage the comparatively knowledge from a richer source domain to a sparser target domain \cite{hu2018conet, yuan2019darec,ni2018perceive}. 


A series of works utilize share-bottom architecture to extracts the auxiliary information from source domains to assist target domain task learning.
CoNet~\cite{hu2018conet} tries to fuse collaborative filtering with deep learning. It shares user features in the embedding layer and enables dual knowledge transfer across domains through cross-mapping. ACDN \cite{liu2019deep} proposes a heterogeneous variants of CoNet by introducing user aesthetic preference. To capture the characteristics and the commonalities of each domain, \citet{ShengZZDDLYLZDZ2021STAR} propose STAR and introduce one centered network shared by all domains and domain-specific networks for each domain. 

Moreover, several works introduce transfer learning methods to further alleviate negative transfer issues~\cite{chen2021user}. 
For instance, DARec~\cite{yuan2019darec} employs a probabilistic framework that leverages matrix factorization to model the rating problem and transfer knowledge across different domains. 
Later, DDTCDR \cite{li2020ddtcdr} utilizes autoencoder to encode user information and item's metadata from online platform, and then adopts latent orthogonal mapping to extract user preferences over multiple domains.


However, most of the cross-domain methods exploit share-bottom architecture so that the training process of source domain and target domain are coupled which makes it unsuitable for super-domain modeling which contains hundreds of billions of data.

\subsection{Pre-Training for Recommendation}
Pre-training approaches are proposed to extract valuable knowledge from large-scale data for specific downstream tasks ~\cite{erhan2009difficulty,DevlinCLT19Bert, zhou2020s3}. 
In the field of recommendation, pre-training is introduced to leverage the side-information to enrich users' and items' representation~\cite{chen2019behavior, charoenkwan2021bert4bitter, zhang2021causerec, wu2020ptum}.

Inspired by the great success of BERT~\cite{DevlinCLT19Bert} in Natural Language Processing, BERT4Rec~\cite{sun2019bert4rec} introduces the self-supervised pre-training manner into recommendation. It employs deep bidirectional self-attention to model user behavior sequences and use Cloze~\cite{DevlinCLT19Bert} objective loss.
To overcome the issue that fine-tuning is inefficient in re-training for each new downstream task, PeterRec~\cite{yuan2020parameter} allows the pre-trained parameters to remain unaltered during fine-tuning by injecting a series of re-learned neural networks. 
UPRec~\cite{xiao2021uprec} leverages the user attributes and
structured social graphs to construct self-supervised objectives in the pre-training stage for extracting the heterogeneous user information. For capturing the deeper association between context data and sequence data, S3-Rec~\cite{zhou2020s3} utilizes the intrinsic data correlation. It can derive self-supervision signals and devise four auxiliary self-supervised objectives via utilizing the mutual information maximization.

Pre-training based methods can utilize the huge amount of data in source domain. However, the fine-tuning mechanism is usually adopted in these methods which may face the catastrophic forgetting problem in continuously-trained industrial recommender systems. In this paper, we propose a knowledge extraction and plugging framework to tackle this issue.

\section{Preliminaries}
\label{sec:preliminaries} 
\subsection{Deep User Response Prediction Model} 
\label{sec:pre_1}
User response prediction (e.g., CTR, CVR and add-to-cart prediction) tasks play an important role in many industrial systems. 
In the era of deep learning, the most common modeling architecture of above tasks usually consists of three components -- the embedding layer, the feature interaction module and the MLP (Multi-Layer Perception) layer which are illustrated by Figure ~\ref{fig:overall} (c). 

Formally, the embedding layer takes the corresponding user features and item features as input and maps them into a vector of dense values. 
Then, we take such vectors of dense values as the input of feature interaction module such as attention mechanism\cite{zhou2018din, wu2022feedrec}, GRU\cite{zhou2019dien}, co-action unit\cite{bian2020can}, etc. Feature interaction module aims to model complex feature relation and output the interaction representation. 
Finally, the resulting representation further passes through several MLP layers (we define $h_{i}$ as the output of $i$-th MLP layer) for final prediction. 

\subsection{Online Learning for Industrial Recommender System} 
\label{sec:pre_2}

\begin{figure}[t]
    \centering
    \includegraphics[width=0.9\columnwidth]{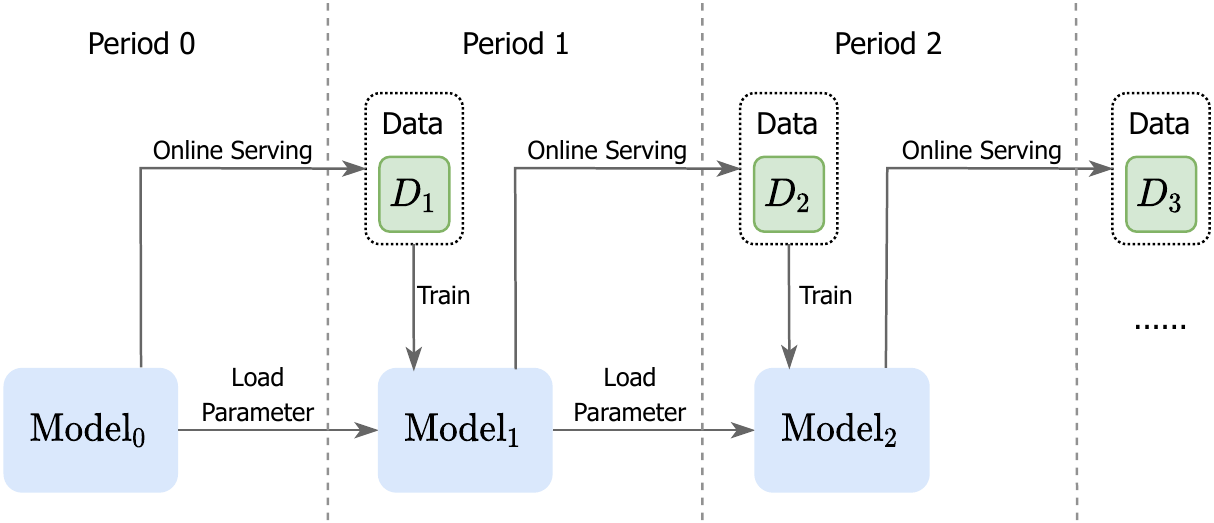} 
    \caption{An illustration of the online learning process for industrial recommender systems.} 
    \label{fig:OL}
\end{figure}

As mentioned in Section~\ref{sec:intro}, to address the data distribution shift issue, industrial recommender systems usually refresh their prediction models in a timely manner, often in the daily or even hourly basis~\cite{sahoo2018online,zhang2020retrain,GuSFZZ2021Defer}. In practice, the training data is collected from real-time traffic log, and then divided by the training frequency. As illustrated by Figure~\ref{fig:OL}, during each training period, the recommendation model will load model parameters from the last version and consume the latest data samples. Such a training process may continue for a long period of time, even for years, in industrial systems like the displaying advertisement system of Alibaba.

Such an online learning strategy makes it easy to accommodate model parameters with the evolving user interests; however, it also brings in other challenges. 
In practice, such a continuously-trained model often requires a significant amount of time to adapt if it can not load all parameters from the previous version which is well-trained by a long-time optimization. 
Thus, it is challenging to recover the performance of model when the model can not load whole well-trained parameters because of changed architecture\footnote{Even if there are only trivial changes such as adding a new feature embedding which changes the dimension of the MLP layer. }. 
With regard to this issue, we introduce a knowledge plugging module for incorporating the pre-trained knowledge from the super-domain without any changes of the model architecture.

\section{Methodology} 
\label{sec:model} 

\begin{figure*}[ht] 
    \centering 
    \includegraphics[width=0.815\textwidth]{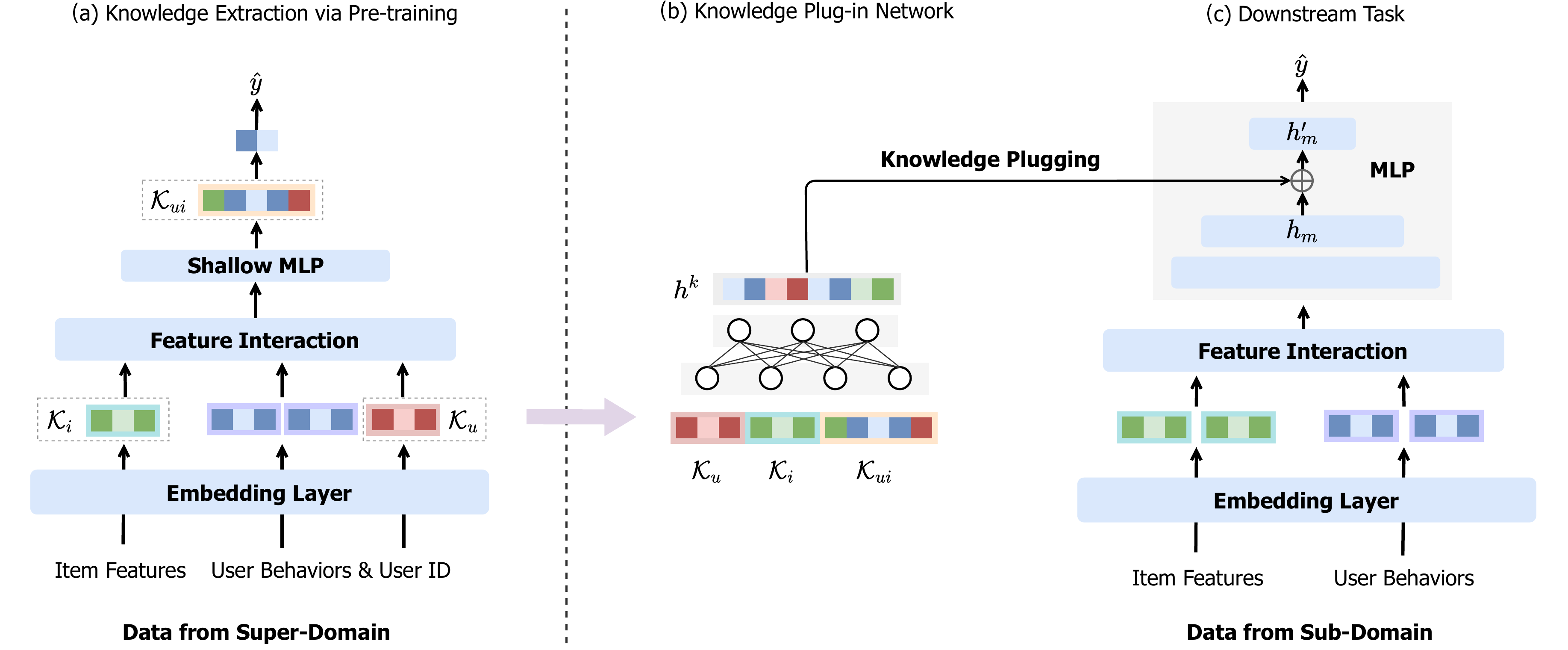} 
    \caption{An illustration of the overall architecture for deep CTR prediction (Figure c), and KEEP which consists of two stages. In the first stage (Figure a), knowledge about user interest is extracted from super-domain in a supervised pre-training manner. Afterward, in the second stage (Figure b), such knowledge is fused with the main click prediction task using a plug-in network.} 
    \label{fig:overall} 
\end{figure*} 


\subsection{Overall Workflow}
As shown in Figure~\ref{fig:overall}, KEEP consists of two stages, namely the the knowledge extraction stage and the knowledge plugging stage. 

In the first stage, we extract knowledge from the super-domain through a knowledge extraction module (see Figure~\ref{fig:overall}(a)). Unlike the common self-supervised pre-training for Natural Language Processing and Computer Vision, we adopt a simple yet effective pre-training approach in the supervised manner (see more details in Section~\ref{sec:ke}). With such a pre-trained model, we utilize the embedding of some specific features, and the output of the second last MLP layer as the extracted knowledge from the super domain. Meanwhile, in order to consume such a large amount of data in the super domain more effectively, we design a concise model architecture for knowledge extraction.

During the second stage, we incorporate the extracted knowledge into the downstream task using the sub-domain data. Different from the conventional fine-tuning mechanism, we take the extracted knowledge as auxiliary input of downstream tasks rather than parameter initialization to avoid catastrophic forgetting in continuously-trained recommender systems. 
As shown in Figure~\ref{fig:overall} (b), a plug-in network is designed to incorporate extracted knowledge without changing downstream model architecture, so that the downstream model can inherit parameters from a previously well-trained model which is friendly with industrial online learning systems (details in Section~\ref{sec:kp}). 


\subsection{Knowledge Extraction}
\label{sec:ke}

We design a supervised pre-training approach for knowledge extraction module because of the simplicity of obtaining labeled feedback data in the recommender systems.
More specifically, a large-scale user feedback data (e.g., click) is firstly collected, and on top of that, a user feedback prediction task is formulated for pre-training. 
In this work, we take into account three tasks for pre-training, including the click prediction, the conversion prediction (whether to purchase a product or not) and the add-to-cart prediction. 
We argue that each of them encodes a different kind of knowledge for user interests. 
It is worth noting that the click prediction is trained over the impression data whereas the other two tasks are trained on top of user clicks.

\subsubsection{Model Architecture} 
\label{sec:pretrain-model}
As illustrated by Figure~\ref{fig:overall} (a), the knowledge extraction module utilizes the same model architecture as DIN~\cite{zhou2018din}. 
The input contains both item features and user features.
Due to the sheer volume of data in the super domain, we adopt a concise model architecture with shallow MLP layers and simplified features~\footnote{including user id, user behavior sequence, item id, shop id and category id.} for training efficiency.
Besides, the user identifier information is included as a feature for pre-training, which is not considered in our main task as such information is sparse during the training of the main task.

To be more precise, an embedding layer firstly maps all of the input features to the high-dimensional space. Then, the attention mechanism is applied for learning the interaction between item features and user features. Meanwhile, DIN compresses all of the embedding, along with the feature interaction, into to one vector of dense values. In the end, a four-layer MLP with the last layer exporting a 2-dimensional dense values is exploited for final output.

\subsubsection{Model Details}
\begin{figure}[ht]
    \centering
    \includegraphics[width=0.825\columnwidth]{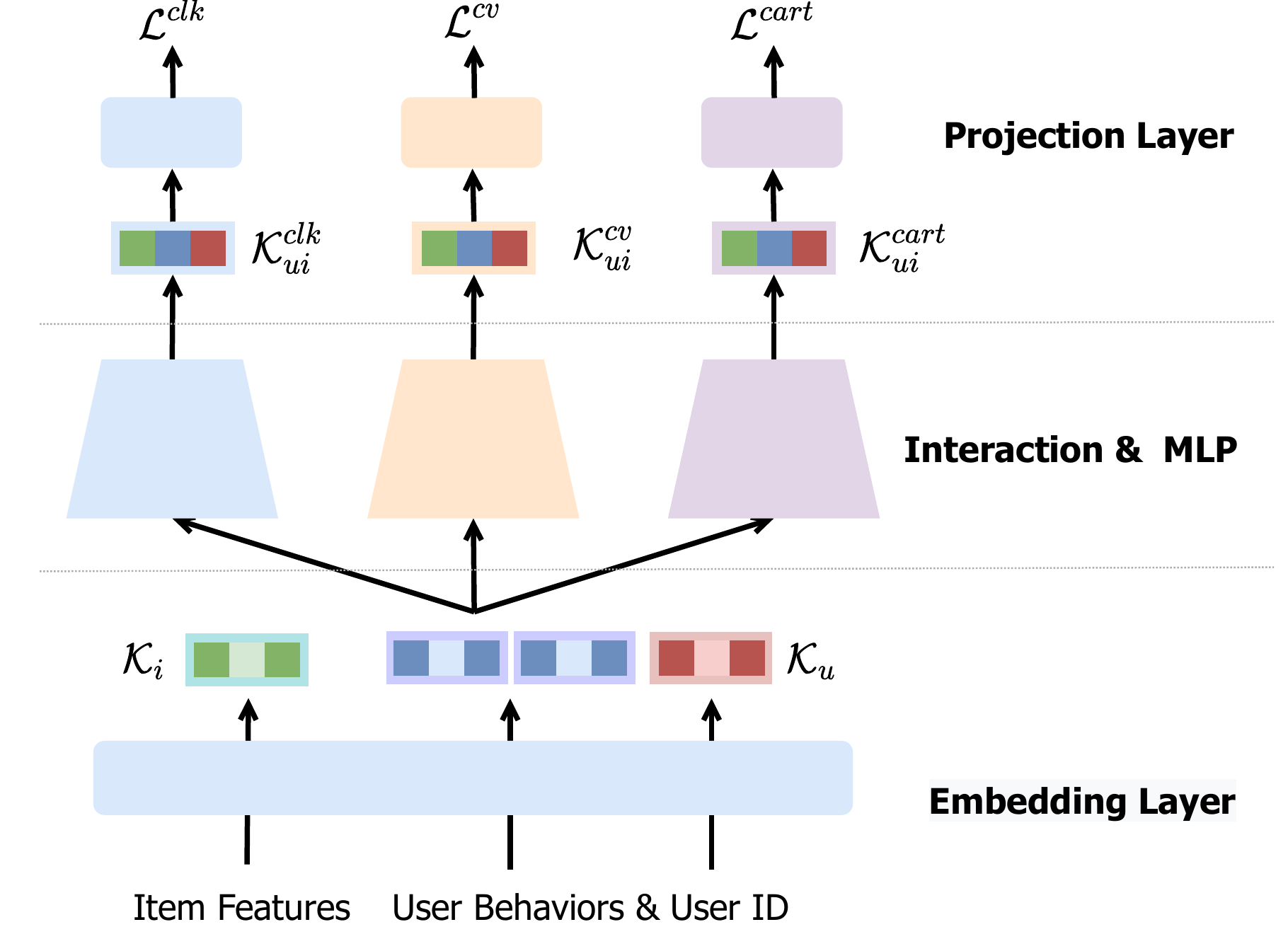} 
    \caption{An illustration of multi-task based supervised pretraining for KEEP. $\mathcal{L}^{clk}$, $\mathcal{L}^{cv}$ and $\mathcal{L}^{cart}$ denote the loss functions for click prediction, conversion prediction and add-to-cart prediction. 
    $\mathcal{K}_u$, $\mathcal{K}_i$ and $\mathcal{K}_{ui}$ stand for the extracted user-level, item-level and user-item interaction knowledge.
    } 
    \label{fig:MTL}
\end{figure}

KEEP is pre-trained in the multi-task manner with three tasks on predicting click, conversion and add-to-cart behaviors, respectively. As shown in Figure ~\ref{fig:MTL}, these three tasks are jointly trained with the share-bottom model structure, in which the embedding layer is shared across multiple tasks while each task has its private MLP layers. Note that the pre-training objectives are not limited to the above three tasks, and the modeling structures (e.g., feature interaction and MLP layers) can also be further improved. However, the main focus of this paper is to validate the efficacy of the overall framework, whereas further advancements for each component are left for future exploration. In the below sections, we will discuss the model implementation details of our training tasks.

\begin{equation}
\begin{aligned}
\label{eq:pretrain-ctr}
\small
\mathcal{L}^{point} = &\sum_{(u, i, c)\in\mathcal{D}} -c \log \sigma(s_{ui}) - (1-c) \log (1-\sigma(s_{ui}))
\end{aligned} 
\end{equation}

Click prediction is conducted over the dataset $\mathcal{D}$ containing user impressions and clicks. For each user-item pair ($u$, $i$), our goal is to optimize the cross-entropy loss between the predicted click probability and the binary click label $c$, as shown in Equation~\ref{eq:pretrain-ctr}. Here, $s_{ui}$ stands for the logit of model output, which can be converted to the click probability after sigmoid transformation. The conversion prediction and the add-to-cart prediction can be defined in a similar way except the dataset is constructed over all of user clicks.

Due to the data sparsity of user feedback (conversion and add-to-cart in particular), we also employ the pairwise optimization approach following~\citet{li2015click}. Specifically, instead of considering each user-item pair ($u$, $i$) independently, we introduce an additional item $j$ with a different feedback label to construct a triplet ($u$, $i$, $j$)~\footnote{In the most common cases, item $i$ and item $j$ are selected from the same session when constructing those triplets.}. The pairwise loss, as illustrated by Equation~\ref{eq:pairwise-loss}, then attempts to maximize the likelihood of separating user-item pairs with different labels. Again, $s_{ui}$ and $s_{uj}$ denote the logits of model output. One potential benefit for such an approach, as mentioned in ~\citet{li2015click}, is that through the pairwise comparison, we could obtain much more supervised signals during model training. In this paper, we only take into account the pairwise logistic loss for simplicity. We believe that other forms of pairwise loss can also be exploited, where we leave them for future exploration~\cite{pasumarthi2019tf}.

\begin{equation}
\begin{aligned}
\label{eq:pairwise-loss}
\small
\mathcal{L}^{pair} 
     = \sum_{(u, i, j)\in\mathcal{D}} \log (1 + \exp(-(s_{ui} - s_{uj})))
\end{aligned} 
\end{equation} 

A pure pairwise loss cannot yield calibrated model predictions, that is, the predicted output does not align with the click probability. For this reason, a hybrid model that combines both pairwise loss and pointwise cross-entropy loss can be adopted~\cite{li2015click}. Therefore, we apply the hybrid loss as shown in Equation~\ref{eq:hybrid}, with the hyper-parameter $\alpha$ tuning for the weight of pairwise loss. Note that this approach can be easily adapted for any of the three tasks.

\begin{equation}
\begin{aligned}
\label{eq:hybrid}
\small
\mathcal{L}= \mathcal{L}^{point} + \alpha \mathcal{L}^{pair}
\end{aligned} 
\end{equation} 

\subsubsection{Output Representation} 
\label{sec:pretrain-output}
As illustrated by Figure~\ref{fig:overall} (c), we categorize the extracted knowledge into three groups: user-level knowledge $\mathcal{K}_u$, item-level knowledge $\mathcal{K}_i$ and user-item interaction knowledge $\mathcal{K}_{ui}$. In this work, we represent $\mathcal{K}_u$ with the embedding of user id, and $\mathcal{K}_i$ with the embedding of item features. In addition, $\mathcal{K}_{ui}$ is represented with the dense vector produced by the second last MLP layer (i.e., the layer before the logit output layer).

The adoption of shared-bottom based multi-task model further results in three different kinds of user-item interaction knowledge, including $\mathcal{K}_{ui}^{clk}$, $\mathcal{K}_{ui}^{cv}$ and $\mathcal{K}_{ui}^{cart}$ that are extracted from the click prediction, conversion prediction and add-to-cart prediction tasks, respectively. Note that since the embedding layer is shared across multiple learning objectives, the user-level and item-level knowledge stay the same for different prediction tasks. In the end, all of the extracted knowledge will be concatenated to construct the final output $\mathcal{K}(u,i)$ for the main downstream task. 

\begin{equation}
\label{eq:k}
\mathcal{K}(u,i) = [\mathcal{K}_{u}, \mathcal{K}_{i}, \mathcal{K}_{ui}^{clk}, \mathcal{K}_{ui}^{cv}, \mathcal{K}_{ui}^{cart}]
\end{equation}

\subsection{Knowledge Plugging} 
\label{sec:kp}
After obtaining the pre-trained knowledge, our next step is to incorporate such knowledge into the downstream task (such as CTR prediction). 
The simplest way is to treat it as an additional feature, which is often achieved by concatenating $\mathcal{K}(u, i)$ with the output $h_m$ of the m-th MLP layer. However, this introduces a big challenge for online-learning based recommendation service that is widely adopted in industrial systems. Online learning systems prefer a stable model architecture to avoid training partial model parameters and feature interactions from scratch. To this end, we utilize a \textbf{Plug-in Network} that can fuse the extracted knowledge with unchanged model architecture. 

Plug-in network replaces feature concatenation with adding operation, i.e., adding the $\mathcal{K}(u, i)$ to $h_m$. 
To better align the dimension size between $h_m$ and $\mathcal{K}(u, i)$, we exploit a shallow MLP structure for dimension projection. More precisely, the extracted knowledge $\mathcal{K}(u, i)$ is firstly fed into the projection layer, and the resulting dense vector $h^k$ will be added with $h_m$. In practice, we discover that the addition operation is equivalent to the concatenation of an extra feature, whereas the former requires less time for fine-tuning in an online learning based recommendation service. The above process can be visualized as Figure \ref{fig:overall}(b). Note that such a network structure is easily extensible if the additional knowledge is available, where we only need to project the new knowledge with another knowledge plug-in network and add it into $h'_m$.

\begin{equation}
\begin{aligned}
\label{eq:plug-in} 
h^k &= \textbf{MLP}(\mathcal{K}(u, i)) \\
h'_m &= h_m + h^k
\end{aligned}
\end{equation}

Hereto, we accomplish the extracted knowledge plugging by the plug-in network. The useful extracted knowledge can assist the performance in the downstream task.

\begin{figure*}[ht]
    \centering
    \includegraphics[width=0.925\textwidth]{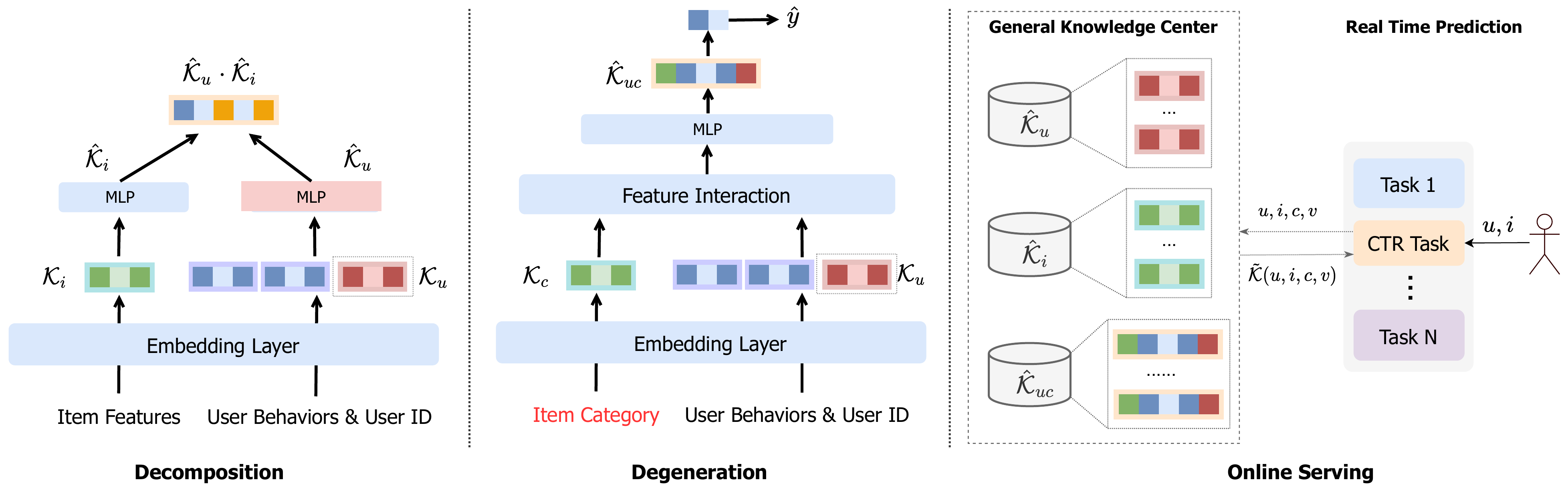} 
    \caption{An illustration of online serving for the proposed GKC. GKC follows the decomposition and degeneration strategies to cache the extracted knowledge in a high performance parameter server. In practice, GKC receives a specific request of a user $u$, a candidate item $i$, its corresponding category $c$, and a version indicator $v$, and then returns the cached knowledge $\tilde{\mathcal{K}}(u,i,c,v)$.} 
    \label{fig:online_serving}
\end{figure*}

\section{Industrial Design For Online Serving} 
\label{sec:empirical}
In this section, we introduce our hands-on experience on implementing KEEP for the display advertising system in Alibaba.

\subsection{Challenges}
Online model inference in Alibaba display advertising system is achieved by a Real-Time Prediction (RTP) System~\cite{Pi2019MIMN}. After receiving each user request, RTP firstly extracts necessary model features and ranks candidates with the predicted probability. Latency is a critical measure for RTP as the system usually requires processing a massive number of user requests within a short time. 

This imposes a great challenge for KEEP as it introduces extra computation for knowledge extraction. Particularly, the user-item interaction knowledge $\mathcal{K}_{ui}$ is obtained through a feature interaction module that may lead to a dramatic growth of inference latency. Despite that the model architecture can be largely reduced, its memory usage and the shallow MLP structure still add extra burden for online serving. To this end, we adopt an empirical strategy by caching the extracted knowledge in a high-performance parameter server to avoid real-time computation. 

A direct caching of all user-item specific knowledge $\mathcal{K}(u,i)$ is a great engineering challenge since the number of user-item pairs is a multiplication of the number of users and the number of items. In the display advertising platform of Alibaba, we have billions of users, and over millions of advertisements, making it impossible to cache all of the pairs. For this reason, we propose to experiment with two strategies: decomposition strategy and degeneration strategy.

\subsection{Decomposition and Degeneration for KEEP} 
\label{sec:dcdg}
\subsubsection{Knowledge Extraction Decomposition} 
The decomposition strategy attempts to replace the current knowledge extraction model with a two-tower structure~\cite{covington2016deep}. We decompose user-item interaction knowledge $\mathcal{K}_{ui}$ into the element-wise product of user-level knowledge $\hat{\mathcal{K}}_u$ and item-level knowledge $\mathcal{\hat{K}}_i$. As illustrated by Figure~\ref{fig:online_serving}(a), with this architecture, we only need to store the user-level and item-level knowledge, which greatly reduce the amount of entries for the caching system.

\subsubsection{Knowledge Extraction Degeneration} 
Despite being efficient, the decomposition strategy scarifies the interaction between user features and item features that often leads to sub-optimal model performance~\cite{zhu2018learning}. Therefore, as shown in Figure~\ref{fig:online_serving}(b), we further propose a degeneration strategy that still keeps the feature interaction while the item-related features (e.g., the identifier of an item) are replaced by category-related features. In this way, $\mathcal{\hat{K}}_{uc}$ (here, $c$ denotes the category of an item) can be viewed as an alternative to $\mathcal{\hat{K}}_{ui}$. Since the number of categories (less than 1,000 in our production system) is much fewer than the number of items, the degenerate strategy significantly reduces the number of entries required for caching system. 

Overall, the knowledge decomposition and degeneration strategies can be visualized with Figure~\ref{fig:online_serving}, and we further summarize its overall process in the below Equation~\ref{eq:kuc}. 
\begin{align}
\label{eq:kuc}
\mathcal{K}(u,i)\rightarrow \mathcal{\hat{K}}(u,i,c) = [\mathcal{\hat{K}}_u,  \mathcal{\hat{K}}_i, \mathcal{\hat{K}}_u \cdot \mathcal{\hat{K}}_i, \mathcal{\hat{K}}_{uc}]
\end{align} 
With the decomposition and degeneration strategy, the number of entries required for knowledge caching can be reduced dramatically from $\mathcal{N}_u * \mathcal{N}_i$ to $\mathcal{N}_u + \mathcal{N}_i + \mathcal{N}_{uc}$, in which $\mathcal{N}_u$, $\mathcal{N}_i$ and $\mathcal{N}_{uc}$ indicate the total number of users, items and user-category pairs, respectively. 

\subsection{GKC: General Knowledge Center}

Industrial recommendation service often consists of a number of subsystems, serving for multiple scenes and multiple prediction tasks. For instance, in the display advertising platform of Alibaba, there exists more than 50 online models at the same time. To facilitate the reuse of the extracted knowledge, we develop a separate General Knowledge Center (GKC) module which can be easily adopted by any online prediction models. 
GKC caches the extracted knowledge in a high-performance parameter server and responses corresponding cached knowledge to downstream tasks when receives request. 

As mentioned in Section~\ref{sec:pre_2}, production models often need to be updated in periodically. KEEP decouples the modeling of super-domain and sub-domain into two stages so that the pre-training model (using the super-domain data) and downstream model (using the sub-domain data) can be trained separately and updated in the different time periods. Since the extracted knowledge is part of model input, to guarantee the consistency between offline training and online inference, we also need to develop a proper synchronization strategy for GKC.

Unfortunately, this is extremely difficult as the online prediction model and GKC are deployed in different systems. To deal with this issue, GKC maintains multiple versions of extracted knowledge at the same time. A downstream model may request the extracted knowledge from GKC according to its own version during model training through a version indicator $v$.
More specifically, we first extend a user-item pair ($u$, $i$) into a user-item-category-version quadruple ($u$, $i$, $c$, $v$), and during online serving, GKC receives a batch of such quadruples and returns the cached knowledge $\mathcal{\tilde{K}}(u,i,c,v)$. In practice, we limit the maximum number of stored versions to 5, which is sufficient, in our production system, to guarantee the consistency in practice. Such a solution requires four times more storage than caching only one version, however, we find that it is necessary for the stability of online serving.

\section{Experiments}
\label{sec:exp}
This section starts with introducing the adopted dataset for experimentation, which is a short-term eight-day impression log sampled from Taobao homepage recommender system. With such a dataset, we conduct an extensive number of experiments by comparing KEEP with a variety of baseline approaches. Finally, the effectiveness of KEEP is further justified through online A/B experiment on the display advertising system of Alibaba.

\subsection{Dataset}
\label{sec:data sets}
Due to the lack of a large-scale super-domain dataset for industrial recommender system, we construct experimental data with the impression log from our production system. This dataset consists of two parts: a super-domain dataset and a sub-domain dataset. More precisely, the super-domain dataset collects the impression log of Taobao homepage recommender system during a eight-day period of time (from 2021/12/13 to 2021/12/20). It includes the recommended results from multiple subsystems, such as the recommended products, recommended videos and recommended advertisements\footnote{In Taobao APP, videos and advertisements are also associated with sailing products.}. The sub-domain dataset only records the impression log of the online advertising recommender system from 2021/12/18 to 2021/12/20.

The overall statistics of the above datasets are provided in Table~\ref{table:Statistics}. It can be found that the super-domain contains 27 times more impressions than the sub-domain, whereas the number of users is only twice as much. This helps alleviate the data sparsity for each user. In addition, such data will be divided into training and testing depending on the task settings. As mentioned in the Introduction section, KEEP is closely related to the below two tasks.

\begin{itemize}[leftmargin=*] 
    \item[$\bullet$] The \textbf{cross-domain recommendation} task aims to improve the target-domain model by utilizing the data of a source domain. To align with this task setting, we select the super-domain data from 2021/12/18 to 2021/12/19 as the source domain, and the sub-domain data from the same time period as the target domain. Such data will be used for training, while the sub-domain data in 2021/12/20 will be used for testing.
    
    \item[$\bullet$] 
    The \textbf{pre-training based recommendation} task requires a large scale data for pre-training, and often needs to be fine-tuned with a certain amount of data for better adapting to the downstream tasks. To this end, we adopt the super-domain data from 2021/12/13 to 2021/12/17 for pre-training, and the sub-domain data from 2021/12/18 to 2021/12/19 for fine-tuning. Again, the sub-domain data in 2021/12/20 will be adopted for testing. 
\end{itemize}

It is worth noting that user behaviors such as click, conversion and add-to-cart are all recorded for labels. During training, the data will be used only once (i.e. training for one epoch), and will be consumed in the manner of incremental training. That is, data samples are organized in the ascending order of date, and are shuffled within each day. Model performance will be reported on the testing set after the training process is completed.
\vspace{-1mm}
\subsection{Compared Methods}
To evaluate the effectiveness of KEEP, we compare it with several 
strong 
algorithms on cross-domain recommendation and pre-training based recommendation. Overall, we take into account two vanilla baselines (\textbf{CAN} and \textbf{Sample Merging}), two cross-domain recommendation algorithms (\textbf{CoNet} and \textbf{DARec}) and two pre-training based recommendation models (\textbf{BERT4Rec} and \textbf{S3-Rec}). 

\begin{table}[]
\caption{The overall statistics of the super-domain dataset and sub-domain dataset.}
\vspace{-3mm} 
\footnotesize 
\centering
\begin{tabular}{lcccccc}
\toprule
& impression & click & conversion & cart & user \\ \midrule
super-domain & 5.8e10 & 1.9e9 & 3.4e7 & 2.4e7 & 2.3e8 \\ 
sub-domain   & 2.1e9  & 8.2e7 & 9.1e5 & 8.3e5 & 1.2e8 \\
\bottomrule
\end{tabular}
\label{table:Statistics}
\end{table} 

\begin{itemize}[leftmargin=*] 
\item {\textbf{Base.}} CAN~\cite{bian2020can} is adopted as the base approach since it is the production model serving for the main traffic of Alibaba displaying advertisement platform. 

\item{\textbf{Sample Merging.}} The introducing of extra data itself may bring in significant performance gain. We thus experiment with a simple consolidation of the feedback data from both super domain and sub-domain for model training.

\item{\textbf{CoNet~\cite{hu2018conet}}} 
introducing cross connections between the source network and the target network to facilitate the dual knowledge transfer across multiple domains. 

\item{\textbf{DARec~\cite{yuan2019darec}}} is a deep domain adaption model for cross-domain recommendation. It is built on top of the idea of DANN~\cite{ganin2015unsupervised} with slight modifications for task accommodation. 

\item{\textbf{BERT4Rec~\cite{sun2019bert4rec}}} adopts the transformer-based modeling structure of BERT~\cite{DevlinCLT19Bert}, and constructs a pre-training task to predict masked items of users' click sequences.

\item{\textbf{S3-Rec~\cite{zhou2020s3}}} devises four self-supervised tasks for learning better feature representation by maximizing mutual information.

\item{\textbf{KEEP}} is our proposed approach. It firstly conducts knowledge extraction through pre-training on the five-day super-domain data. The extracted knowledge is then fine-tuned with two-day sub-domain data for a better accommodation of the sub-domain. Comparing to the cross-domain recommendation setup, it employs extra five-day super-domain data. To this end, we introduce a new variant of KEEP by excluding such data, and we name it as \textbf{KEEP-C}. KEEP-C keeps the knowledge extraction and plugging process unchanged except the pre-training and plugging are co-trained with the same data. 
\end{itemize}

\textbf{Implementation details.} 
As mentioned in Section~\ref{sec:ke}, we adopt concise model architecture and select only a small subset of features to accelerate the training process in the knowledge extraction stage. 
Embedding of those features are shared across multiple pre-training tasks and the embedding dimension is set to 48 for user id and 16 for all of the other features. The MLP layers are set to [512, 256, 128, 64, 2], and each pre-training task owns its private MLP parameters. For each task, the output of the 4-th layer is adopted to represent user-item-interaction knowledge. 

The extracted knowledge will then be incorporated to a downstream task for better recommendation quality. Our downstream model is aligned to the production deployment, which includes several effective components such as target attention, search-based interest modeling~\cite{PiZZWRFZG2020SIM} and Co-Action Unit~\cite{bian2020can}. Both of the pre-training and downstream models are trained with Adam with learning rate 0.001 and the batch size is 2000. Besides, the loss weight $\alpha$ in Equation~\ref{eq:hybrid} is set to 0.25. All experiments are built on XDL2~\cite{zhang2022picasso}. 

\textbf{Evaluation Metric.} 
The ranking performance is evaluated with the Group AUC (\textbf{GAUC}) which is the top-line metric in our production system and has shown to be more consistent with the online performance~\cite{ZhuJTPZLG2017OCPC,zhou2018din,ShengZZDDLYLZDZ2021STAR}. 
GAUC can be computed with Equation~\ref{eqn:gauc}, 
\begin{equation}
\label{eqn:gauc} 
\begin{aligned} 
    \textrm{GAUC} = \frac{\sum_{u=1}^U \# \textrm{impressions}(u) \times \textrm{AUC}_u}{\sum_{u=1}^U \# \textrm{impressions}(u)}
\end{aligned}
\end{equation}
in which $U$ represents the number of users, $\#\textrm{impression}(u)$ denotes the number of impressions for the $u$-th user, and $\textrm{AUC}_u$ is the AUC computed using the samples from the $u$-th user. 
\textbf{In practice, a lift of 0.1\% GAUC metric (in absolute value difference) often results in an increase of 1\% CTR.} 
Hereafter, we will ONLY report the absolute value difference for GAUC. 

\subsection{Experimental Results}
\subsubsection{Overall Performance} 
Table~\ref{tab:result_offline} offers a comparison of GAUC performance over different baseline approaches. Here, our task is to make a better prediction of user click, i.e. CTR prediction. We discover that a simple merge of data samples (i.e., Sample Merging) leads to a significant drop of model performance, meaning that a proper data fusion mechanism needs to be devised for better utilization of the extra data. Both of the cross-domain and pre-training based recommendation algorithms provide such a mechanism, and as shown in Table~\ref{tab:result_offline}, they all outperform the Base model.

Compared to the baseline approaches, KEEP achieves the best GAUC performance, outperforming the Base method by 0.7\%, and the best baseline method by 0.3\%. This demonstrates the effectiveness of our proposed knowledge extraction-and-plugging paradigm. To understand the pure effect of the extraction-and-plugging structure, we introduce KEEP-C that excludes the five-day data for pre-training. This offers exactly the same experiment setup as the cross-domain recommendation. Results from Table~\ref{tab:result_offline} show that KEEP-C still exhibits a marginal improvement over CoNet and DARec. It again validates our proposed modeling structure.

\begin{table}
\centering
\footnotesize 
\caption{A comparison of GAUC performance across different baseline methods on CTR prediction. The standard deviation is computed but not provided because the value is smaller than 0.0002. (We are tested with billions of samples.)}
\vspace{-2mm}
\label{tab:result_offline}
\begin{tabular}{lcc}
\toprule
 Compared Methods        & \quad\quad GAUC \quad\quad\quad  &   Improv. \\ 
\midrule
{\quad Base}~ & $0.6310 $ & - \\ 
{\quad Sample Merging}~    & $ 0.6185 $  & $-0.0124$   \\ \midrule 

\multicolumn{3}{l}{Cross-domain Recommendation} \\ 
{\quad CoNet}~    & $ 0.6336 $  & $+0.0027$  \\
{\quad DARec}~    & $ 0.6341 $  & $+0.0032$  \\
{\quad KEEP-C}~    & $0.6348$  & $+0.0039$  \\ \midrule

\multicolumn{3}{l}{Pre-training based Recommendation} \\ 
{\quad BERT4Rec}~    & $ 0.6326 $  & $+0.0017$  \\
{\quad S3-Rec}~    & $ 0.6341 $  & $+0.0030$  \\ \midrule
{\quad KEEP}~    & $0.6380$  & $+0.0070$  \\
\bottomrule
\end{tabular}
\end{table}

\subsubsection{The Effect of $\mathcal{K}_u$, $\mathcal{K}_i$ and $\mathcal{K}_{ui}$.} As mentioned in Figure~\ref{fig:overall}, we extract three types of knowledge during pre-training. To understand the effectiveness of each one, we design the below ablation study. Table~\ref{tab:result_ablation1_offline} lists the GAUC performance for a set of combinations of knowledge utilization. An adoption of user-level knowledge $\mathcal{K}_u$ results in a lift of GAUC, and a further introducing of item-level knowledge $\mathcal{K}_i$ provides an additional gain for modeling performance. A final consolidation of all three types of extracted knowledge yields the best performance, proving the effectiveness of utilizing the extracted knowledge in KEEP.

\begin{table}
\centering
\footnotesize 
\caption{The effect of different types of extracted knowledge $\mathcal{K}_u$, $\mathcal{K}_i$ and $\mathcal{K}_{ui}$. Note that we take into account three pre-training tasks (click, conversion and add-to-cart) when extracting the user-item interaction knowledge $\mathcal{K}_{ui}$.}
\vspace{-2mm}
\begin{tabular}{ccc|cc}
    \toprule
    $\mathcal{K}_{u}$ &$\mathcal{K}_{i}$ &$\mathcal{K}_{ui}$ & GAUC \quad &  Improv. \\
    \midrule 
                  &              &               & $0.6310$ & -         \\ 
    $\checkmark$  &              &               & $0.6332$ & $+0.0022$ \\ 
    $\checkmark$  & $\checkmark$ &               & $0.6347$ & $+0.0037$ \\ 
    $\checkmark$  & $\checkmark$ &  $\checkmark$ & $0.6380$ & $+0.0070$ \\ 
    \bottomrule
\end{tabular}
\label{tab:result_ablation1_offline}
\end{table}

\subsubsection{The Effect of Pre-training Tasks.} 
We also conduct an ablation study to examine the utility of different pre-training tasks. Here, our goal remains to be the incorporating of pre-trained knowledge on downstream CTR prediction. As illustrated by Table~\ref{tab:result_ablation2_offline}, pre-trained knowledge from the click prediction task helps boost the GAUC by a relatively large margin. By incorporating the pre-trained knowledge from conversion and add-to-cart prediction tasks, we see a further improvement of the GAUC performance. 

Moreover, pre-trained knowledge purely from the click prediction task brings the best performance improvement, which might be due to two reasons. First, click is the most common behavior in a recommender system so that it has a rich amount of data available for training. Second, since our downstream task is to predict user click, a pre-training on the click task will be a better fit. 

\begin{table}
\centering
\footnotesize 
\caption{The effect of utilizing the extracted knowledge from three different types of pre-training task.}
\vspace{-3mm}
\begin{tabular}{ccc|cc}
\addlinespace
    \toprule
    Click & Conversion & Add-to-Cart & GAUC \quad &  Improv. \\
    \midrule 
                 &               &              & $0.6310$ & -          \\
    $\checkmark$ &               &              & $0.6368$ & $+0.0058$  \\
    $\checkmark$ &  $\checkmark$ &              & $0.6374$ & $+0.0064$  \\
    $\checkmark$ &  $\checkmark$ & $\checkmark$ & $0.6380$ & $+0.0070$  \\
    \bottomrule
\end{tabular}
\label{tab:result_ablation2_offline}
\end{table}

\subsection{Production Deployment}

Furthermore, we evaluate KEEP through production deployment in the display advertisement system of Alibaba. 
For production deployment, we exploit an extremely large-scale dataset with almost two-year of impression log from the super-domain for knowledge extraction. In addition, a 45-day impression log from the online advertising system, along with the extracted knowledge, will be used for training our production models. Again, those models are deployed on the display advertisement platform of Alibaba. 

\subsubsection{Performance on Different Downstream Tasks.} We first analyze the performance of KEEP in an offline manner, that is, the trained production models will be tested with an additional one-day impression log. To understand the utility of KEEP on different downstream tasks, we report GAUC for CTR prediction, CVR prediction and add-to-cart prediction, which are the most common tasks in e-commence recommender systems. As shown in Table~\ref{tab:production_multi_task}, KEEP improves the downstream CTR, CVR and add-to-cart prediction tasks by 0.0042, 0.0036 and 0.0065 on the GAUC performance, respectively. This demonstrates the utility of the extracted super-domain knowledge across multiple domains, and the generalizability of our proposed extraction-and-plugging modeling framework.

\begin{table}
\centering
\footnotesize 
\caption{The GAUC performance of KEEP in three different downstream tasks with the trained production models.}
\vspace{-3mm}
\begin{tabular}{lcc|c}
    \addlinespace
    \toprule
    Prediction Task      & Base  & KEEP &  Improv. \\
    \midrule        
    {CTR}~ & $0.6661$ & $0.6703 $ & $+0.0042$ \\
    {CVR}~    & $0.7513$ & $0.7549$  & $+0.0036$   \\
    {CART}~    & $0.6286$ & $ 0.6351 $  & $+0.0065$  \\
    \bottomrule
\end{tabular}
\label{tab:production_multi_task}
\end{table}

\subsubsection{The Amount of Pre-training Data.} One important factor that affects the pre-training quality is the amount of data. Here, we conduct three experiments that utilizes 1-month, 6-month and 2-year of super-domain data for pre-training, and the resulting performance is provided in Table~\ref{tab:production_amount_data}. We observe a trend of performance growth with an increase of data amount. 

\begin{table}
\centering
\footnotesize 
\caption{The effect of model prediction performance with the amount of pre-training data.}
\vspace{-3mm}
\begin{tabular}{l c | c}
    \addlinespace
    \toprule
    Model        &  GAUC \quad   &  Improv.  \\
    \midrule        
    {Base}~ & $ 0.6661 $ & - \\
    {KEEP (1 month)}~    & $ 0.6667 $  & $+0.0006$   \\
    {KEEP (6 months)}~    & $ 0.6679 $  & $+0.0018$  \\
    {KEEP (2 years)}~    & $ 0.6703 $  & $+0.0042$  \\
    \bottomrule
\end{tabular}
\label{tab:production_amount_data}
\end{table}

\subsubsection{Performance on Different User Groups.} We hypothesize that KEEP can better characterize user interest for long-tailed users.  We thus split them into sub-groups according to their behavior frequency.  More specifically, users with the number of clicks in the range of [0, 50), [50, 150), [150, 300) and 300+ are assigned to different sub-groups.  According to Table~\ref{tab:production_user}, KEEP produces a better performance lift for users whose behaviors are more sparse -- a 0.0054 GAUC lift for the group [0, 50) whereas a 0.0031 lift for the group 300+.  This validates our hypothesis, and KEEP indeed relieves the data sparsity by leveraging the super-domain data.

\begin{table}
\centering
\footnotesize 
\caption{The performance of KEEP on different user groups.}
\label{tab:public}
\vspace{-3mm}
\begin{tabular}{lccc}
    \addlinespace
    \toprule
    User Behaviors & Base & KEEP &  Improv. \\
    \midrule
    {0-50}~   & $0.6573 $  & $0.6627 $  & $+0.0054$ \\
    {50-150}~ & $0.6628 $ & $ 0.6668 $  & $+0.0040$   \\
    {150-300}~& $0.6669 $ & $ 0.6703 $  & $+0.0034$  \\
    {>300}~   & $0.6687 $ & $ 0.6718 $  & $+0.0031$  \\
    \bottomrule
\end{tabular}
\label{tab:production_user}
\end{table}

\subsubsection{Decomposition and Degeneration Strategy.} 
The above sections provide a complete offline evaluation for production model performance, whereas such a model cannot be directly served online. As mentioned in Section~\ref{sec:dcdg}, KEEP proposes a decomposition strategy and a degeneration strategy for industrial deployment. For both of the two strategies, we expect the drop of GAUC and our goal is to minimize the performance loss. Table~\ref{tab:production_ind} shows the GAUC performance of the decomposition and degeneration strategies. The two strategies, when adopted alone, produce a non-trivial drop of performance. However, a combination of both strategies results in only a marginal performance drop. This is the version we adopted in our online production model.

\begin{table}
\centering
\footnotesize 
\caption{The GAUC performance regarding to the decomposition and degeneration strategies for online serving.}
\vspace{-3mm}
\begin{tabular}{l c | c}
    \addlinespace
    \toprule
    Model        & GAUC &  Improv. \\
    \midrule        
    {KEEP}~                              & $ 0.6703 $  & - \\
    {$\text{KEEP}_{decomposition}$}~                   & $ 0.6691 $  & $-0.0012$   \\
    {$\text{KEEP}_{degeneration}$}~                  & $ 0.6683 $  & $-0.0020$  \\
    {$\text{KEEP}_{decomposition} + \text{KEEP}_{degeneration}$}~    & $ 0.6698 $  & $-0.0005$  \\
    \bottomrule
\end{tabular} 
\label{tab:production_ind}
\end{table}

\subsubsection{Online A/B Testing.} With the above online serving strategies, KEEP have been deployed in the display advertising system of Alibaba. During the time period from October 7 to November 14 2021, we conduct a strict online A/B testing to validate the effectiveness of KEEP. 
Compared to our production baseline, \textbf{KEEP achieves a 5.4\% increase of CTR and a 4.7\% increase of RPM (Return Per Mille).} 
Here, both CTR and RPM are the top-line metrics for our production system. 
Now, KEEP has been deployed fully online and serves the main traffic in Alibaba advertisement system. 


\section{Conclusion}
In this work, we propose a novel KEEP framework to extract knowledge from super-domain, and apply such knowledge for online recommendation. KEEP is a two-stage framework that consists of pre-training knowledge extraction on super-domain, and incorporating the extracted knowledge into the downstream model by a plug-in network. For the purpose of production deployment, we further develop the decomposition and degeneration strategies which significantly reduce the online inference latency.

Experiments conducted on an experimental dataset, sub-sampled from the production impression log, demonstrate KEEP outperforms several state-of-the-art cross-domain recommendation and pre-training based recommendation algorithms. It is worth noting that KEEP has been deployed in the display advertising system in Alibaba, bringing a lift of $+5.4\%$ on CTR and $+4.7\%$ on RPM. With the promising results of such proposed supervised multi-task pre-training approaches, we will further explore the pre-training task design, and develop more effective methods to handle the version inconsistency issue of our General Knowledge Center. 

\section*{Acknowledgements}
We would like to thank the anonymous reviewers for their constructive comments. 
This work is supported by Alibaba Group through Alibaba Research Intern Program. 

\balance

\bibliographystyle{ACM-Reference-Format}
\bibliography{sample-base}

\end{document}